\documentclass[12pt,epsfig,a4]{article} 
\newcommand{\vs}{\vspace{-0.25cm}}
\usepackage{amsmath,graphicx,wasysym}
\usepackage{float}
\begin{document} 
\begin{center}
{\Large{\bf Two-pion exchange nucleon-nucleon potentials with Roper resonance excitation} }  

\bigskip

 Norbert Kaiser \\
\medskip
{\small Physik-Department, Technische Universit\"{a}t M\"{u}nchen,
   D-85748 Garching, Germany\\

\smallskip

{\it email: nkaiser@ph.tum.de}}
\end{center}
\medskip
\begin{abstract} Motivated by the recent paper arXiv:2602.11815 [nucl-th], we calculate in these notes the spectral functions (i.e. imaginary parts) of the NN-potentials as they arise from $2\pi$-exchange with single and double Roper resonance excitation. In contrast to the full one-loop calculation, the spectral functions of the isoscalar and isovector central and tensor potentials  can be given in simple analytical form. The pertinent momentum-space potentials $V_C(q), W_C(q), V_T(q), W_T(q)$ are obtained via subtracted dispersion relations. This representation allows also to include a regulator function, that tames high-momentum components of the chiral $2\pi$-exchange. The calculation is extended to $2\pi$-exchange with combined $\Delta(1232)$-isobar and Roper resonance excitation.       
\end{abstract}

\section{Preparation}
In the nonrelativistic treatment the nucleon-nucleon interaction as it arises from two-pion exchange takes the following form:   
\begin{equation} T_{N\!N} = V_C(q) + \vec \tau_1\!\cdot\! \vec \tau_2\, W_C(q) -\big[  V_T(q) + \vec \tau_1\!\cdot\! \vec \tau_2\, W_T(q)\Big] (\vec \sigma_1\!\times\! \vec q\,)\!\cdot\! (\vec \sigma_2\!\times \!\vec q\,)\,,\end{equation}
where $\vec \tau_{1,2}$ and  $\vec \sigma_{1,2}$ denote the isospin- and spin-operators of both nucleons and $\vec q$ is the momentum transfer with magnitude $q = |\vec q\,|$. 
The sign-convention for $T_{N\!N}$ has been chosen such that one-pion exchange gives $T_{N\!N}^{(1\pi)} = (g_A/2f_\pi)^2  \vec \tau_1\!\cdot\! \vec \tau_2\,\vec \sigma_1\!\cdot\! \vec q\,\vec \sigma_2\!\cdot \!\vec q\,(m_\pi^2+q^2)^{-1}$, with $g_A \simeq 1.3$, $f_\pi = 92$\,MeV (pion decay constant), and $m_\pi = 138\,$MeV the average pion mass. The long-range parts of the $2\pi$-exchange NN-potentials are obtained from the corresponding spectral functions (i.e. imaginary parts) via twice- or once-subtracted dispersion relations as:
\begin{equation} V_C(q) = C_0+ C_1\, q^2 +{2q^4 \over \pi} \!\int_{2m_\pi}^\infty \!\!d\mu\, {{\rm Im}V_C(i\mu) \over \mu^3 (\mu^2+q^2)}\,, \qquad V_T(q) = C_T-{2q^2 \over \pi} \!\int_{2m_\pi}^\infty \!\!d\mu\, {{\rm Im}V_T(i\mu) \over \mu (\mu^2+q^2)} \,. \end{equation} 
For the isovector central and tensor potentials $W_C(q),  W_T(q)$ analogous dispersion relations hold.  As a further modification, that is often used, this representation allows to introduce a regulator function, e.g. $f_\Lambda(\mu) = \exp(-\mu^2/\Lambda^2)$, in the spectral integral to tame high-momentum components of the chiral $2\pi$-exchange interaction. 

The imaginary part  of a one-loop $2\pi$-exchange diagram (see Fig.\,1) is calculated with the help of the Cutkosky cutting rule as \cite{strohm}:
\begin{equation}
{\rm Im}\int\!{d^4 l \over(2\pi)^4i}\, {A\otimes B \over (m_\pi^2 - l\!\cdot\!l)(m_\pi^2 -(q\!-\!l)\!\cdot\!(q\!-\!l)) } = {k\over 16 \pi \mu} \int_{-1}^1 \! dx\,  A\otimes B\,, \end{equation}
where $A$ and $B$ symbolize vertices on the left and right hand side of the diagram and the angular integral $\int_{-1}^1\!dx$ effectively goes over the Lorentz-invariant $2\pi$-phase space. Furthermore, $\mu = \sqrt{q\!\cdot\! q}\geq 2m_\pi$ denotes the $2\pi$-invariant mass and $k = \sqrt{\mu^2/4-m_\pi^2}$ is the (single) pion center-of-mass momentum. For the $\bar N\!N\to \pi\pi\to \bar N \!N $ $t$-channel kinematics under consideration in the heavy baryon limit, one sets the nucleon velocity four-vector $v^\nu$ equal to $v^\nu = (0, i\, \vec v\,)$ with $\vec v $ a 3-component unit-vector, which  leads to the relations $v\!\cdot\!q=0$ and $v\!\cdot\!l = -i \,x\, k$. In the first step the emerging integrand is complex-valued, but it becomes real-valued after symmetrizing it with respect to $x\to -x$, and leaves an integral $\int_0^1\!dx$ over positive values of the cosine $x$. The intermediate state nucleon is handled with an infinitesimally small positive mass-splitting $\eta$, which leads to $\lim_{\eta\to 0} \eta/(\eta^2+k^2x^2) = \pi \delta(x)/k$ and $\int_0^1\! dx\, \delta(x)=1/2$.  Further technical details on the method can be found in section\,2.2 of ref.\cite{strohm}, but note that therein the numerator of eq.(9) must be corrected to $\sqrt{\mu^2-4m_\pi^2}$. 

The $\pi NN^*$ vertex and heavy Roper propapator pertinent to the present calculation  read \cite{lisheng}:
\begin{equation} {g'_A\over 2f_\pi} \, \vec \sigma\!\cdot\! \vec l\,\, \tau^a\,, \quad {i \over v\!\cdot \!l -\rho}\,,\qquad\qquad  {3g_A\over 2\sqrt{2}f_\pi} \, \vec S\!\cdot\! \vec l\,\, T^a\,, \quad {i \over v\!\cdot \!l -\Delta}\,, \end{equation} 
with $g'_A$ a coupling constant (various values are discussed in section III.A of ref.\cite{lisheng}) and $\rho \simeq 500\,$MeV the mass difference between the Roper $N^*(1440)$-resonance and the nucleon. The right half of eq.(4) gives the analogous input for $\Delta(1232)$-isobar excitation \cite{meins}, where the $\Delta N$-mass difference is smaller: $\Delta\simeq 300\,$MeV. The $2\times 4 $ spin- and isospin-transition matrices $S_i$ and $T_a$ fulfill the relations $3S_i S_j^\dagger = 2\delta_{ij}-i\epsilon_{ijk} \sigma_k$ and  $3T_a T_b^\dagger = 2\delta_{ab}-i\epsilon_{abc} \tau_c$. The coupling ratio $3/\sqrt{2}$ to $\pi N\!N$ is empirically well satisfied.   
\section{Spectral functions of $2\pi$-exchange NN-potentials}
In this section we present the expressions for the spectral functions Im$V_C(i\mu)$, Im$W_C(i\mu)$, Im$V_T(i\mu)$, Im$W_T(i\mu)$ as they result from the one-loop $2\pi$-exchange diagrams shown in Fig.\,1. We remind of the abbreviation $k$ with its meaning: $2k =\sqrt{\mu^2-4m_\pi^2}$. 
\begin{figure}[H]
\centering
\includegraphics[width=0.9\linewidth]{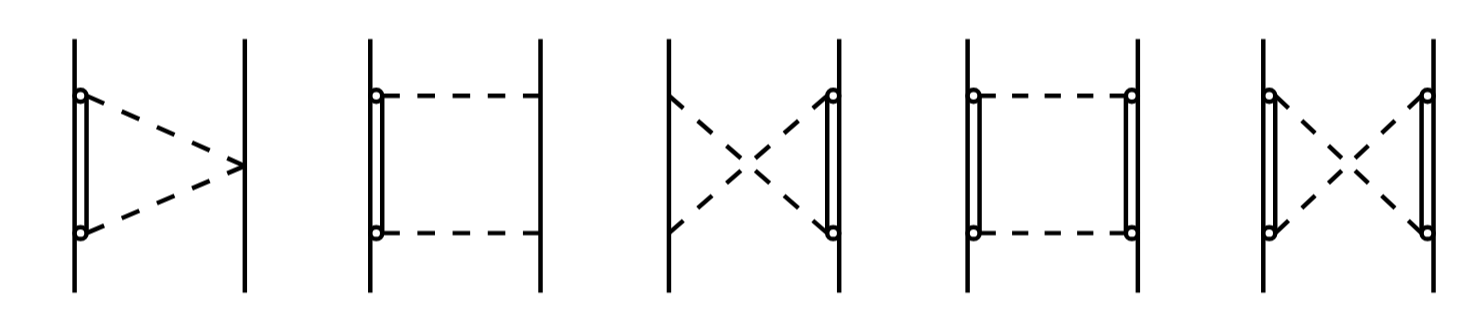}
\caption{One-loop $2 \pi$-exchange diagrams with single and double Roper resonance excitation.}
\end{figure}
\subsection*{Triangle diagrams with Roper excitation}
The (left) triangle diagram in Fig.\,1 and its mirror partner lead to the following (doubled) contribution to the 
imaginary part of the isovector central potential:
\begin{equation} {\rm Im}W_C(i\mu) = {g'^2_A\over \pi \mu (2f_\pi)^4} \bigg\{ \Big({2m_\pi^2 \over 3} -{5\mu^2 \over 12} -\rho^2\Big) k + \rho\Big({\mu^2\over 2}-m_\pi^2+\rho^2\Big) \arctan{k\over \rho}\bigg\}\,. \end{equation} 
Since this diagram involves on one side the isovector  Weinberg-Tomozawa $2\pi N\!N$-contact vertex, there is no $g_A^2$ in the prefactor, as it was incorrectly written in eq.(A1) of ref.\cite{lisheng}.
\subsection*{Planar and crossed box diagrams with single Roper excitation}
The (central) planar and crossed box diagrams in Fig.\,1 with single Roper excitation and their mirror partners lead to the following (doubled) contributions to the spectral functions:
\begin{equation} {\rm Im}V_C(i\mu)=  {3 g_A^2 g'^2_A\over  \mu (4f_\pi)^4 \rho}(\mu^2 -2 m_\pi^2)^2 \,, \qquad  V_C(q) ={6g_A^2 g'^2_A\over \pi (4f_\pi)^4  \rho\, q}(2 m_\pi^2+q^2)^2\arctan{q \over 2m_\pi} \,, \end{equation} 
\begin{equation} {\rm Im}W_C(i\mu) = {g_A^2g'^2_A\over \pi \mu (2f_\pi)^4} \bigg\{ \Big({11\mu^2 \over 12} -{5m_\pi^2 \over 3}+\rho^2\Big) k -{1\over  \rho}\Big({\mu^2\over 2}-m_\pi^2+\rho^2\Big)^2 \arctan{k\over \rho}\bigg\}\,, \end{equation} 
\begin{equation} {\rm Im}V_T(i\mu) = {3g_A^2g'^2_A\over \pi \mu (4f_\pi)^4} \bigg\{{1\over  \rho}(\mu^2-4m_\pi^2+4\rho^2) \arctan{k\over \rho}-4 k \bigg\}\,, \end{equation} 
\begin{equation} {\rm Im}W_T(i\mu)=  { g_A^2 g'^2_A\over  \mu (4f_\pi)^4 \rho}(4m_\pi^2-\mu^2) \,, \qquad  W_T(q) ={2g_A^2 g'^2_A\over \pi (4f_\pi)^4  \rho\, q}(4 m_\pi^2+q^2)\arctan{q \over 2m_\pi} \,. \end{equation}
Modulo additive constants (plus const\,$q^2$) and the opposite sign-convention, the results for $V_C(q)$ and $W_T(q)$ agree with the expressions written in eqs.(5,14) of ref.\cite{lisheng}. 
\subsection*{Planar and crossed box diagrams with double Roper excitation}
The (right) planar and crossed box diagrams in Fig.\,1 with double Roper excitation lead to the following contributions to the spectral functions:
\begin{eqnarray} {\rm Im}V_C(i\mu) &=& {3g'^4_A\over \pi \mu (4f_\pi)^4} \bigg\{ \bigg[2\mu^2 +4\rho^2+{8(\rho^2-m_\pi^2)^2 \over \mu^2 -4m_\pi^2+4\rho^2}\bigg] k \nonumber \\ && +{1\over \rho} \Big( {\mu^2 \over 2} -m_\pi^2 +\rho^2\Big)( \mu^2 -2m_\pi^2 -6\rho^2) \arctan{k\over \rho}\bigg\}\,, \end{eqnarray} 
\begin{eqnarray} {\rm Im}W_C(i\mu) &=& {g'^4_A\over \pi \mu (4f_\pi)^4} \bigg\{ \bigg[{2\over 3}(17\mu^2 -20m_\pi^2)+ 16\rho^2+{16(\rho^2-m_\pi^2)^2 \over \mu^2 -4m_\pi^2+4\rho^2}\bigg] k \nonumber \\ && +{1\over \rho} (2m_\pi^2 -2\rho^2-\mu^2)( \mu^2 -2m_\pi^2 +10\rho^2) \arctan{k\over \rho}\bigg\}\,, \end{eqnarray} 
\begin{equation} {\rm Im}V_T(i\mu) = {3g'^4_A\over \pi \mu (4f_\pi)^4} \bigg\{{1\over  \rho}\Big({\mu^2\over 4}-m_\pi^2+3\rho^2\Big) \arctan{k\over \rho}-3 k \bigg\}\,, \end{equation} \begin{equation} {\rm Im}W_T(i\mu) = {g'^4_A\over \pi \mu (4f_\pi)^4} \bigg\{{1\over  \rho}\Big(2m_\pi^2+2\rho^2-{\mu^2 \over2}\Big) \arctan{k\over \rho}-2 k \bigg\}\,. \end{equation} 
The spectral functions for the $2\pi$-exchange NN-potentials with single and double $\Delta(1232)$-isobar excitation (computed in section\,3 of ref.\cite{meins}) are obtained from the expressions written here in eqs.(5-13) by replacing the coupling constant $g'_A \to g_A$, the mass-splitting $\rho \to \Delta$, and multiplying with weight factors $-1, 2, -1, -1, 1/2, 4, 1, 1, 1/4$, in that order. 
\subsection*{Planar and crossed box diagrams with combined Roper and Delta excitation}
As emphasized in ref.\cite{lisheng}, when including the Roper $N^*(1440)$-resonance one must also consider explicit contributions from $\Delta(1232)$-isobar excitation because of the lower $\Delta N$-mass splitting. The NN-potentials as they arise from single and double $ \Delta(1232)$-excitation have been given in section \,3 of ref.\cite{meins}. Here, we complete the pattern of $2\pi$-exchange with intermediate low-lying nucleon-resonance excitations by considering the combination of the $\Delta(1232)$-isobar and the Roper resonance. The (right) planar and crossed box diagrams in Fig.\,1 with combined $\Delta(1232)$-isobar and Roper excitation and their mirror partners lead to the following (doubled) contributions to the spectral functions:
\begin{eqnarray} {\rm Im}V_C(i\mu) &=& {3g_A^2g'^2_A\over 64\pi \mu f_\pi^4} \bigg\{ 4\rho\Delta\, k +{\rho\over\rho^2\!-\!\Delta^2} (\mu^2 -2m_\pi^2 +2\Delta^2)^2\arctan{k\over \Delta} \nonumber \\ && -{\Delta\over\rho^2\!- \!\Delta^2} (\mu^2 -2m_\pi^2 +2\rho^2)^2\arctan{k\over \rho}\bigg\}\,, \end{eqnarray}
\begin{eqnarray} {\rm Im}W_C(i\mu) &=& {g_A^2g'^2_A\over 64\pi \mu f_\pi^4} \bigg\{\Big[ {1\over 3}(20m_\pi^2-11\mu^2)  -4 \Delta^2-4\rho^2\Big] k \\ &&  -{\Delta\over\rho^2\!-\! \Delta^2} (\mu^2 -2m_\pi^2 +2\Delta^2)^2\arctan{k\over \Delta} +{\rho\over\rho^2\!- \!\Delta^2} (\mu^2 -2m_\pi^2 +2\rho^2)^2\arctan{k\over \rho}\bigg\}\nonumber \,, \end{eqnarray}
\begin{eqnarray} {\rm Im}V_T(i\mu) &=& {3g_A^2g'^2_A\over \pi \mu (4f_\pi)^4} \bigg\{4 k +{\Delta\over\rho^2\!-\! \Delta^2} (\mu^2 -4m_\pi^2 +4\Delta^2)\arctan{k\over \Delta}   \nonumber \\ &&-{\rho\over\rho^2\!- \!\Delta^2} (\mu^2 -4m_\pi^2 +4\rho^2)\arctan{k\over \rho}\bigg\} \,, \end{eqnarray}
\begin{equation} {\rm Im}W_T(i\mu) = {g_A^2g'^2_A\over \pi \mu (4f_\pi)^4} \bigg\{{\Delta\over\rho^2\!- \!\Delta^2} (\mu^2 -4m_\pi^2 +4\rho^2)\arctan{k\over \rho}-{\rho\over\rho^2\!-\! \Delta^2} (\mu^2 -4m_\pi^2 +4\Delta^2)\arctan{k\over \Delta}  \bigg\}\,, \end{equation}
which are obviously symmetric under the permutation of the mass-splittings $\Delta \leftrightarrow \rho$. The apparent complexity of one-loop integrals with heavy baryon propagators involving different mass-splittings boils down on the level of the spectral function to symmetrizing a term involving $\Delta^{-1}\! \arctan(k/\Delta)$ by another term involving $\rho^{-1}\!\arctan(k/\rho)$.

By comparison with the analytical expressions given in section\,3 of ref.\cite{meins}, one learns that the translation of $\mu$-dependent spectral functions into $q$-dependent $2\pi$-exchange NN-potentials proceeds as follows:
\begin{eqnarray} &&  \mu^2 \to -q^2\,, \qquad\quad  {k \over \mu}\to -{L(q) \over \pi} \,, \quad L(q) = {\sqrt{4m_\pi^2+q^2} \over  q} \ln { \sqrt{4m_\pi^2+q^2}+q \over 2m_\pi} \,,\nonumber \\ && {k \over \mu(\mu^2 - 4m_\pi^2 +4 \rho^2)} \to {L(q) - L(2\sqrt{\rho^2-m_\pi^2}\,)\over \pi(4m_\pi^2+q^2-4\rho^2)} \,,  \qquad\quad  {1\over \mu \,\Delta}\arctan{k \over \Delta} \to {2\over \pi} D_\Delta(q)\,, \end{eqnarray}
and an analogous function  $D_\rho(q)$ \cite{meins}, if the mass-splitting $\Delta\simeq 300\,$MeV  is replaced by  $ \rho\simeq 500\,$MeV.

\end{document}